\documentstyle[11pt,moriond,epsfig]{article}

\def\be{\begin{equation}}
\def\ee{\end{equation}}
\def\bea{\begin{eqnarray}}
\def\eea{\end{eqnarray}}

\begin{document}
\begin{flushright}
CDF/PUB/BOTTOM/PUBLIC/5004 \\
FERMILAB-Conf-99/157-E
\end{flushright}
\renewcommand{\thefootnote}{\fnsymbol{footnote}}
\vspace*{4cm}
\title{Two Recent Results on B Physics from CDF}

\author{M.P. Schmidt \\ for the CDF Collaboration\footnote{Presented
at the XXXIVth Rencontres de Moriond, Les Arcs 1800, France, 13-20 March 1999.} }

\address{Department of Physics, Yale University\\
New Haven, CT 06520}

\maketitle\abstracts{
Preliminary results from two recent CDF $b$ physics analyses are
presented.  The first obtains
$\sin (2 \beta ) = 0.79 {+0.41 \atop -0.44}$ from a measurement of
the asymmetry in
$B^0, \overline{B^0} \rightarrow J/\psi \, K^0_{\rm S}$ decays, providing
the best direct indication so far that $CP$ invariance is violated
in the $b$ sector.  The second obtains new results on the parity
even ($A_0$ and $A_\parallel$) and odd ($A_\perp$) polarization
amplitudes from full angular analyses of
$B^0 \rightarrow J/\psi \, K^{\ast 0}$
and $B_s^0 \rightarrow J/\psi \, \phi$ decays:
\[
   \begin{array}[c]{c}
      B^0: \left\{
      \begin{array}[c]{rcl}
         A_0 & = & 0.770 \pm 0.039 \pm 0.012 \\
         A_\parallel & = & (0.530 \pm 0.106 \pm 0.034) e^{(2.16 \pm 0.46 \pm
                            0.10) i} \\
         A_\perp & = & (0.355 \pm 0.156 \pm 0.039) e^{(-0.56 \pm 0.53 \pm
                        0.12) i}
      \end{array}
      \right.
   \\
   \\
      B_s^0: \left\{
      \begin{array}[c]{rcl}
         A_0 & = & 0.778 \pm 0.090 \pm 0.012 \nonumber \\
         A_\parallel & = & (0.407 \pm 0.232 \pm 0.034) e^{(1.12 \pm 1.29
                            \pm 0.11) i} \\
         |A_\perp| & = & 0.478 \pm 0.202 \pm 0.040
      \end{array}
      \right.
   \end{array}
\]
}

\newpage
\section{Introduction}

A rich program in $b$ production and decay physics has been pursued with data
collected by CDF in Run I (1992 -- 1995).  By making use of a silicon
strip vertex detector and the copious production of various species of
$b$~hadrons at the Tevatron $\bar{{\rm p}}$p collider, we have
obtained precise measurements of the $B^0$, $B^+$, and $B_s^0$ lifetimes and
the $B_s^0$ mass.
We have observed the decay $B_c \rightarrow J/\psi \, \ell \, \nu$,
and set
the most stringent limits
on the decays $B_{d,s} \rightarrow \mu^+ \, \mu^-$ and
$B \rightarrow K^{(\ast)} \, \mu^+ \, \mu^-$.
We also have obtained
competitive measurements on neutral $B$ mixing and ratios of branching
fractions for selected $b$ hadron decay modes.
Most of these results and others have been published or submitted for
publication.~\cite{ref:web}

Preliminary results from two more analyses have become available this year.
The first analysis builds upon and significantly extends
the previously published CDF result~\cite{ref:kelley} on the $CP$
nonconserving parameter $\sin (2 \beta )$ as determined from the asymmetry in
$B^0, \overline{B^0} \rightarrow J/\psi K^0_{\rm S}$ decays.
The analysis presented here includes events that are not fully
reconstructed in the silicon vertex detector, thereby doubling the
available data sample, and uses the combination of three flavor tagging
algorithms. 

The second analysis
obtains results for the polarization amplitudes in the
pseudoscalar to vector-vector decays
$B^0 \rightarrow J/\psi \, K^{\ast 0}$
and $B_s^0 \rightarrow J/\psi \, \phi$.
These results are relevant to understanding the
decay dynamics of hadrons containing a heavy-quark
and provide information relating to the possible use of these
decays for studies of $CP$ invariance violation.
For example, if the decay $B^0 \rightarrow J/\psi \, K^{\ast 0}$
were to occur in a parity
eigenstate (even or odd) followed by the $CP$ invariant
decay $K^{\ast 0} \rightarrow
K_{\rm S}^0 \, \pi^0$, then this mode
could be used as simply as
the $B^0 \rightarrow J/\psi \, K_{\rm S}^0$ mode
for determining $\sin (2 \beta )$.

\section{Improved Measurement of $\sin (2 \beta)$
with $B^0 \rightarrow J/\psi \, K^0_{\rm S}$ Decays}

It has long been recognized~\cite{ref:sanda} that a measurement
of the $CP$ noninvariant asymmetry

\begin{displaymath}
   A_{CP}(t)= {{dN\over{dt}}(\overline{B^0} \rightarrow J/\psi K^0_{\rm S}) -
               {dN\over{dt}}(B^0 \rightarrow J/\psi K^0_{\rm S})\over {
               {dN\over{dt}}(\overline{B^0} \rightarrow J/\psi K^0_{\rm S}) +
               {dN\over{dt}}(B^0 \rightarrow J/\psi K^0_{\rm S})}} 
\end{displaymath}

\noindent provides a phenomenologically
clean method for determining $\sin (2 \beta)$.
Violation of
$CP$ invariance can arise in the Standard Model through a non-trivial
phase in the CKM quark mixing matrix. Interference between the direct decay,
$B^0 \rightarrow J/\psi \, K^0_{\rm S}$, and the decay after mixing ($\overline{B^0}
\rightarrow B^0$) leads to an asymmetry $A_{CP}(t) =  \sin2\beta \sin(\Delta
m t) $. The $B^0 - \overline{B^0}$ mixing frequency is governed by the
mass difference, $\Delta m$,
between the heavy and light mass eigenstates. The proper time of decay, $t$,
is employed in order to achieve maximum sensitivity via a time-dependent
measurement of the asymmetry.  Previous direct measurements of the
asymmetry have been made by OPAL~\cite{ref:opal} and CDF,~\cite{ref:kelley}
the latter result being updated by the analysis presented here.

The CDF measurement of $\sin (2 \beta)$ is made possible by the distinctive
decay to a final state with 
all charged particles: $B^0 \rightarrow J/\psi \, K^0_{\rm S}
 \rightarrow \mu^+ \, \mu^- \, \pi^+ \, \pi^-$.
Charged particle three-momenta are determined at CDF with
an 84-layer drift chamber (the CTC) that covers the
pseudorapidity interval $|\eta| < 1.1$, where
$\eta = - ln[\tan (\theta /2) ] $ and $\theta$ is the polar angle
angle
in a cylindrical coordinate system in which the $z$ axis coincides
with the $\bar{{\rm p}}$p beam line.  The $z$ coordinate
of the $\bar{{\rm p}}$p
interaction is determined with a
time projection drift chamber (the VTX) located inside
the CTC.  The VTX surrounds the silicon vertex detector
(the SVX) which consists
of four layers of axial
silicon strips (providing $r - \phi$ information)
located at radii between 2.9 and 7.9 cm and extending $\pm 25$ cm in $z$ from
the detector center. 
The central tracking volume is immersed in a 1.4 T uniform axial
magnetic field.  
The component of charged track momentum transverse to the beam line, $p_T$, is
determined with a resolution of
$\delta p_T / p_T = [(0.0009 \; \cdot p_T^2({\rm GeV}/c)^2
+ (0.0066)^2]^{1/2}$ for tracks well measured in the SVX-CTC.

Electrons and muons are readily distinguished within
CDF from other charged particles
(pions, etc.) 
The central tracking volume is surrounded by calorimetry with
projective tower geometry
which is augmented with a
preshower detector and with strip chambers at electromagnetic shower
maximum.  Electrons are identified by their interactions in the calorimeter
and by $dE/dx$ information from the CTC and preshower detectors.  Muons
in the central region with $p_T > 1.4 \; {\rm GeV}/c$
typically penetrate the calorimeter
($\sim 5$ absorption lengths) and are detected in central muon chambers
(85\%
coverage in azimuth for $|\eta | < 0.6$). Additional coverage is provided by
central muon upgrade chambers (80\% azimuthal coverage, after a total of
$\sim 8$ absorption lengths) and central extension muon chambers
(67\% azimuthal coverage for $0.6 < |\eta | < 1.0$, after a total of
$\sim 6$ absorption lengths).

\begin{figure}
   \centerline{\epsfxsize 3.2 truein \epsfbox{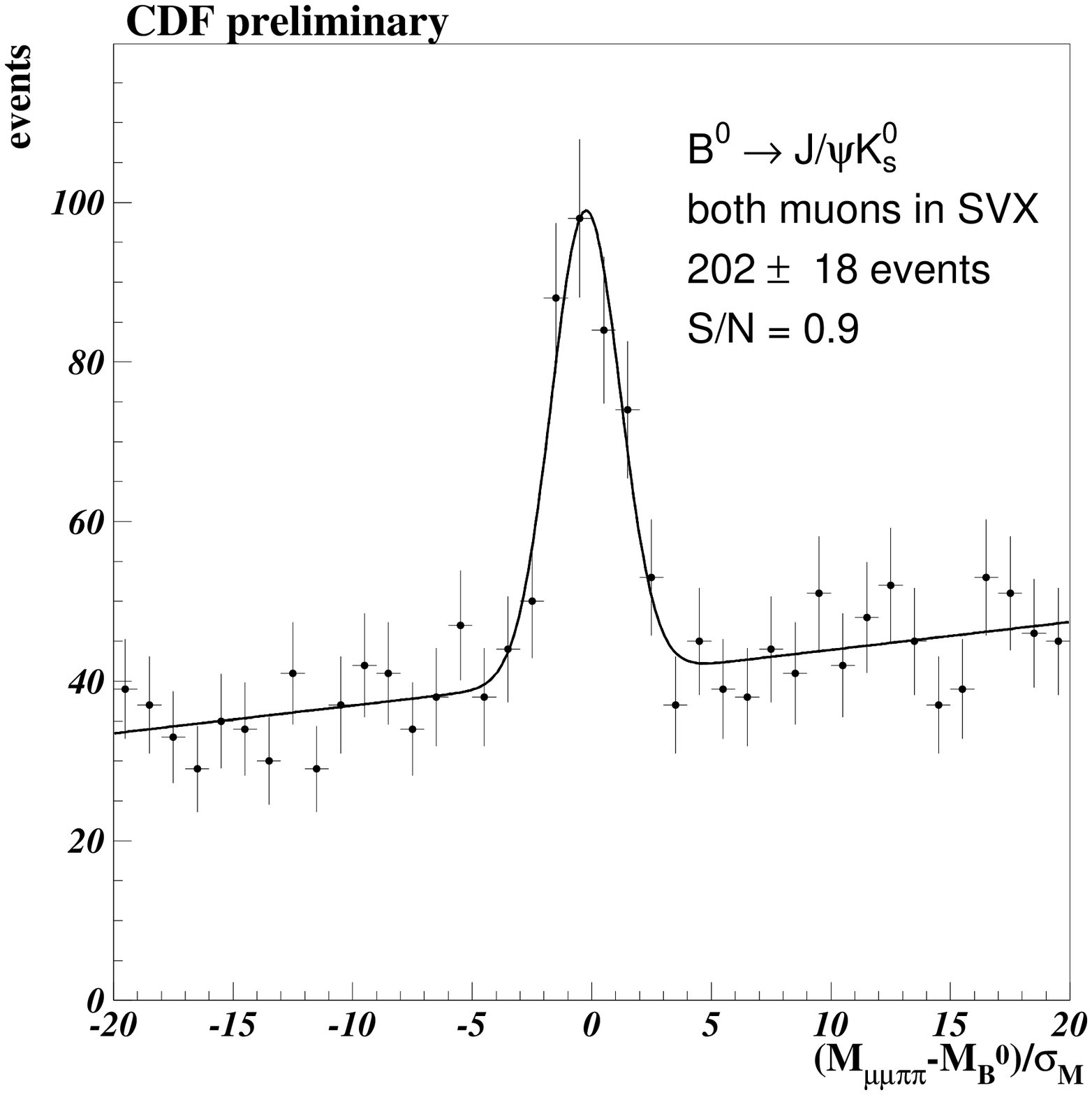}
               \hspace*{0.5cm}
               \epsfxsize 3.2 truein \epsfbox{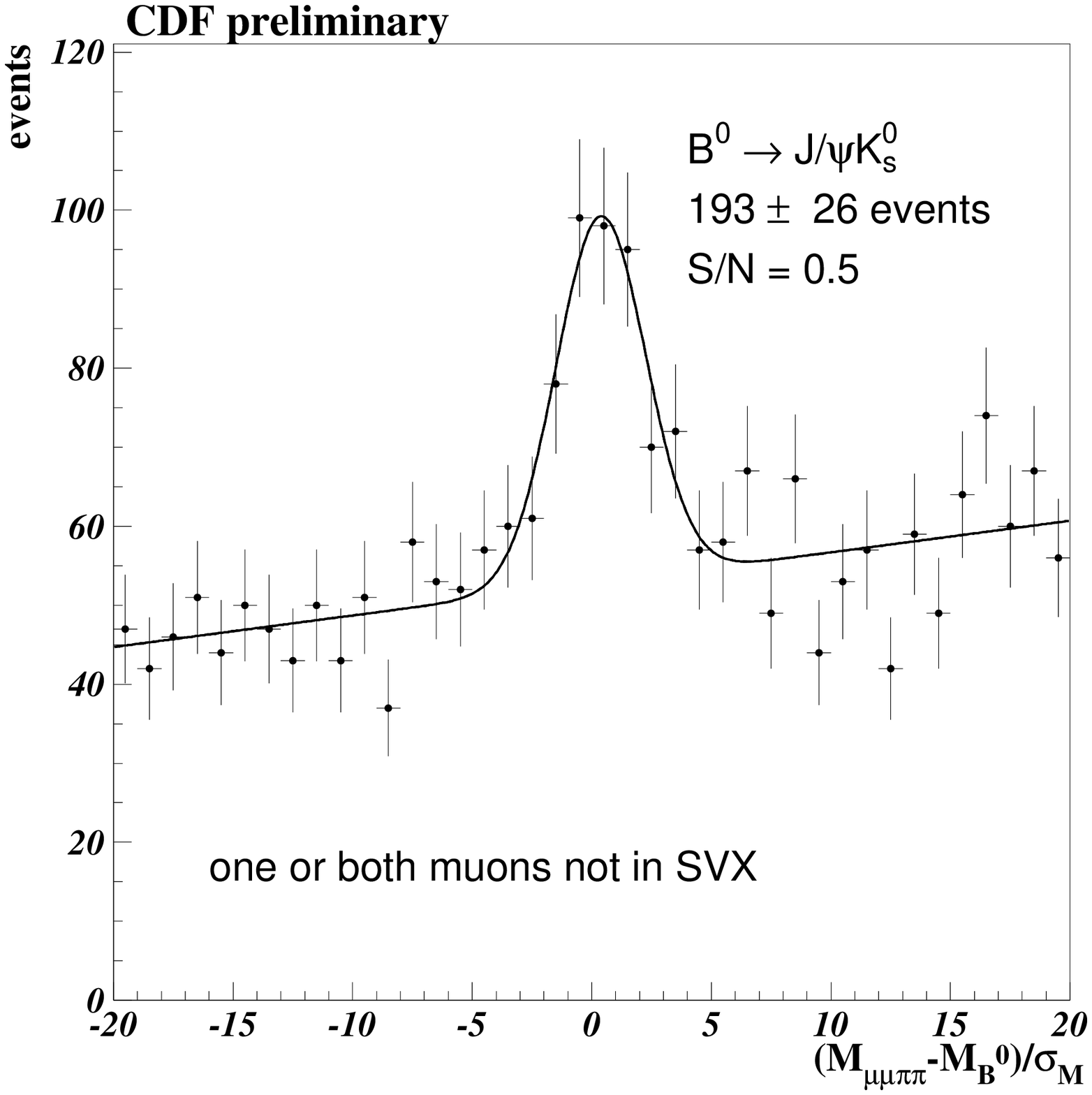}}
   \caption[]
           {\label{fig:norm_mass} Normalized mass distributions for
            $J\psi \, K_{\rm S}^0$ candidates in the SVX and non-SVX samples.}
\end{figure}

Candidate events are reconstructed mainly from data collected with a dimuon
trigger having a relatively low threshold
($p_T (\mu) > \, \sim 2 \; {\rm GeV}/c$).
For events with a reconstructed $J/\psi \rightarrow \mu^+ \, \mu^-$ decay,
$K^0_{\rm S}$ candidates are formed from pairs of oppositely
charged tracks, assumed to be pions, and required to have
$p_T > 0.7 \; {\rm GeV}/c$
and be well separated from the $\bar{\rm p}$p collision envelope.  Mass,
vertex and pointing constraints are then used in a four particle fit
for $B^0$ candidates.
 
From 110 pb$^{-1}$ of data collected with CDF the yield of
$J/\psi \, K^0_{\rm S}$
decays is $395 \pm 31$ events with a signal to background of 0.7 for $p_T (B) >
4.5 \; {\rm GeV}/c$.  The normalized mass distribution is displayed in Fig.
\ref{fig:norm_mass} where the data has been divided into two disjoint
subsets: an SVX sample 
($202 \pm 18$ events with a signal to background of 0.9)
and
a non-SVX sample ($193 \pm 26$ events with a signal to background of 0.5). 
The SVX sample is the subset of candidates for which
both muons had trajectories well measured in the silicon vertex detector. 
The sample sizes are roughly equal due to the limited
acceptance of the SVX (60\%); it's size ($\pm 25 \, {\rm cm}$)
is similar to the $\sim30$ cm rms spread
in the distribution of $\bar{\rm p}$p interactions along the beam axis.

The SVX sample is essentially the same as that employed for
the previously published CDF result~\cite{ref:kelley} on $\sin (2 \beta)$.
The SVX events have the precise decay length information needed
to carry out a time-dependent
asymmetry measurement. The non-SVX events have less precisely determined
decay lengths, but can nevertheless contribute at least via a time-integrated
measurement.  In fact 30\% of the non-SVX events have one muon well measured
in the silicon vertex detector.
 
In order to measure the asymmetry $A_{CP}$ it is necessary to
identify (tag) whether the decaying $B$~meson was initially produced as $B^0$
or $\overline{B^0}$.   The effectiveness of the tag depends on its
efficiency ($\epsilon$) and its purity.  The purity of a tagging algorithm is
usually expressed in terms of a dilution factor
or fractional difference of right
($R$) and wrong ($W$) tags: $D = (N_R - N_W)/(N_R + N_W)$.  An impure tagger
with dilution $D < 1$ results in a
smaller observable asymmetry: $A^{obs}_{CP}
= D A_{CP}$. The statistical uncertainty for the result on $\sin (2 \beta)$
is inversely proportional to $\sqrt {\epsilon D^2}$.

Three tagging algorithms are employed in order to maximize the sensitivity
of the
measurement.  All three methods have been developed and used by CDF for
$B^0 - \overline{B^0}$
mixing measurements.  One of the algorithms employed, a same-side tagging
algorithm (SST), exploits charge correlations expected~\cite{ref:sst} between
the $B$~meson flavor and and the charge of pions produced in fragmentation
or from the decays of resonances ($B^{\ast \ast}$).
The SST algorithm employed with
the SVX sample in the analysis reported here is identical to that used
in the published analysis~\cite{ref:kelley}
on $\sin (2 \beta)$
and an associated mixing measurement.~\cite{ref:petar} 
Appropriate modifications have been made to validate
and apply the SST tag to the
non-SVX sample. The expected dilution and efficiency for the SST algorithm
then largely follows from the previous work.~\cite{ref:kelley,ref:petar}

Two opposite-side tagging algorithms are employed.
A soft lepton tagging algorithm (SLT)
exploits the correlation of the flavor ($b$ or $\bar{b}$) of the $B$~meson at
production with the charge of the lepton from the semileptonic decay of the
(opposite-side) $b$~hadron produced in association with it.
A jet charge tagging
algorithm (JETQ)  exploits a similar correlation between the $B$~meson
flavor
and the momentum weighted sum of charges for a cluster of tracks (a
jet) associated with the decay of the opposite-side $b$~hadron. 
The SLT and JETQ
algorithms are very similar to algorithms used in a CDF mixing
analysis~\cite{ref:sltjetq} carried out with a sample of candidate $B$ mesons
detected via their semileptonic decays.  The $B$ mesons in the mixing
analysis sample have a higher $p_T$ (typically a factor of $\sim 2$)
than the
$J/\psi K^0_S$ events, and this motivates modifications of the SLT and
JETQ algorithms. The expected dilutions and efficiencies for the SLT and JETQ
tagging algorithms are determined with a sample of ~1000 $B^{\pm} \rightarrow
J/\psi \, K^{\pm}$ decays and a sample of ~40,000 inclusive (non-prompt)
$J/\psi \rightarrow \mu^+ \, \mu^-$ decays.  The SLT and JETQ algorithms
are applied to both the SVX and non-SVX event samples.

\begin{table}
\caption[]
{Percentage efficiencies and dilutions as measured for the flavor
tagging algorithms employed.
For the SST algorithm, the efficiency
   values include the fractions of SVX and non-SVX events; thus, the
   tagging efficiency for the total sample is the sum of the SVX and non-SVX
   efficiencies.
}
\begin{center}
\begin{tabular}{ccccc}
\hline \hline
type  &tagger & class & efficiency($\epsilon$)  & dilution$({\cal D})$  \\
\hline \hline
same-side &
same-side & SVX $\mu$ & $35.5\pm 3.7$ &
$16.6\pm2.2$  \\
& same-side & non-SVX $\mu$ & $38.1 \pm 3.9$ &
$17.4 \pm 3.6$ \\  \hline
opposite side &
soft lepton & all events & $5.6\pm 1.8$ &
$62.5 \pm 14.6$ \\
& jet charge & all events & $40.2 \pm 3.9$ & 
$23.5 \pm 6.9$ \\
\hline \hline
\end{tabular}
\end{center}
\end{table}

The tags are defined to be essentially orthogonal; in particular, tracks
within a cone of
$\sqrt{(\eta ^2 + \phi ^2)} < 0.7$ centered on the vector momentum
of the $B \rightarrow
J/\psi K_{\rm S}^0$ decay can be candidates for an SST tag but
are excluded from use for a JETQ tag.
Each event can be tagged by zero, one or two algorithms. In the case of two
tags, one must be SST. If both SLT and JETQ tags are present the SLT
assignment is taken due to its superior (larger) dilution.
Tagging information is
obtained for 80\% of the events in the
$J/\psi K^0_S$ sample.
Taking into account
single and double tags,
a combined effective tagging efficiency
$\epsilon D^2 = 6.3 \pm 1.7$\% is obtained.

An unbinned maximum likelihood~\cite{ref:moresin2beta} fit is used to extract
a value for $\sin (2
\beta)$. The $B^0$ lifetime and $\Delta m_d$ are constrained in the fit to
the world average values. The fit includes the SVX and non-SVX samples and
treats the decay length uncertainty and dilutions appropriately. The fit
allows for prompt and non-prompt background
components as well as the possibility of charge asymmetries in the
efficiencies and dilutions of the tags.  No significant asymmetries are
observed in the dilutions or the backgrounds.

The result from the fit is $\sin (2 \beta ) = 0.79 \pm 0.39 ({\rm stat}) \pm
0.16 ({\rm syst})$.
The statistical uncertainty
dominates and the systematic uncertainty arises almost entirely from the
determination of the dilution parameters with the limited sample of $B^{\pm}$
decays.  
From this result, a 93\% confidence interval of $ 0.0 < \sin (2 \beta ) <
1.0$ is obtained for the frequentist approach advocated by
Feldman and Cousins.~\cite{ref:fandc} Similar limits are obtained
using alternative methods.

\begin{figure}
   \centerline{\epsfxsize 3.2 truein \epsfbox{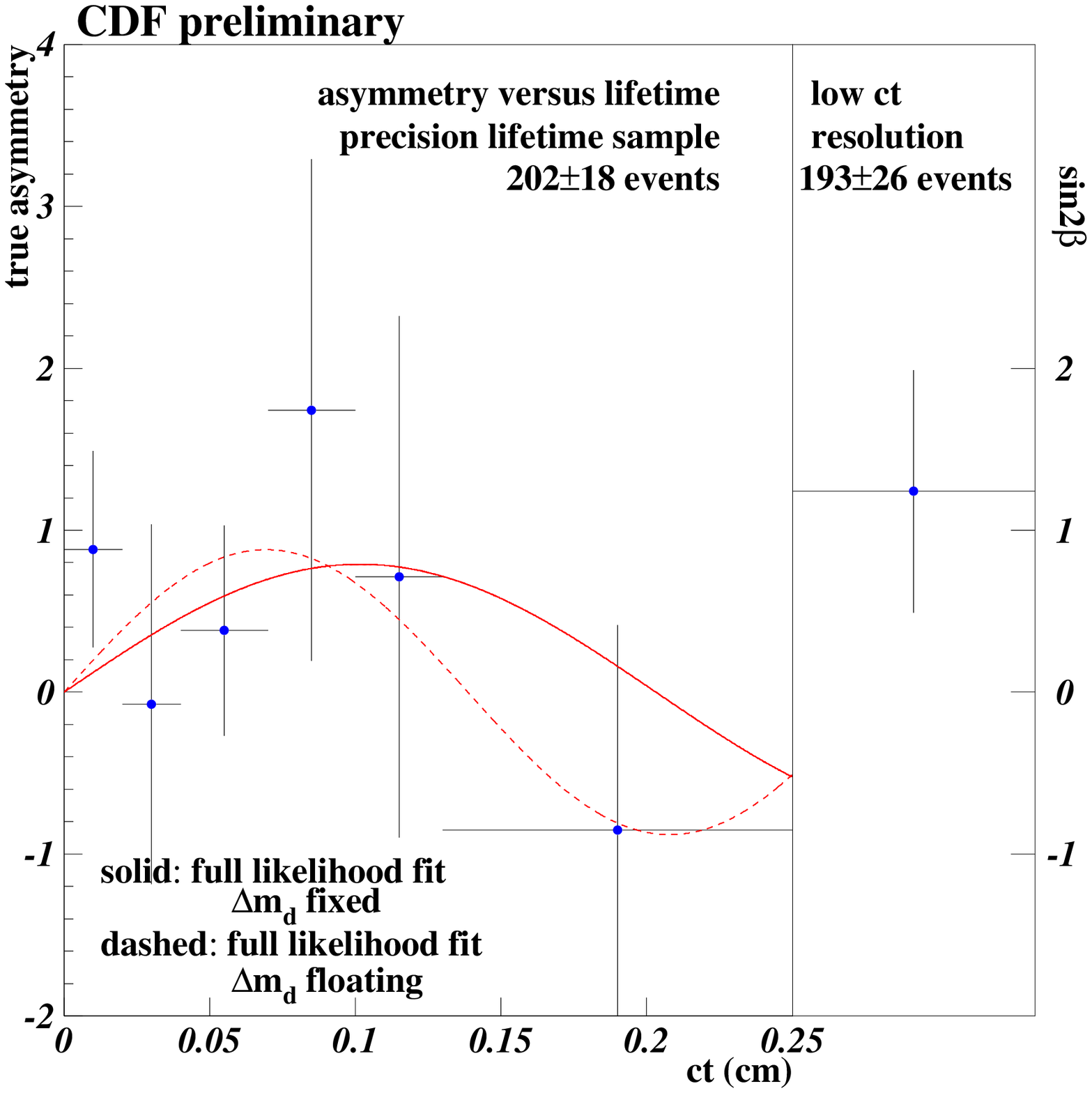}
               \hspace*{0.5cm}
               \epsfxsize 3.2 truein \epsfbox{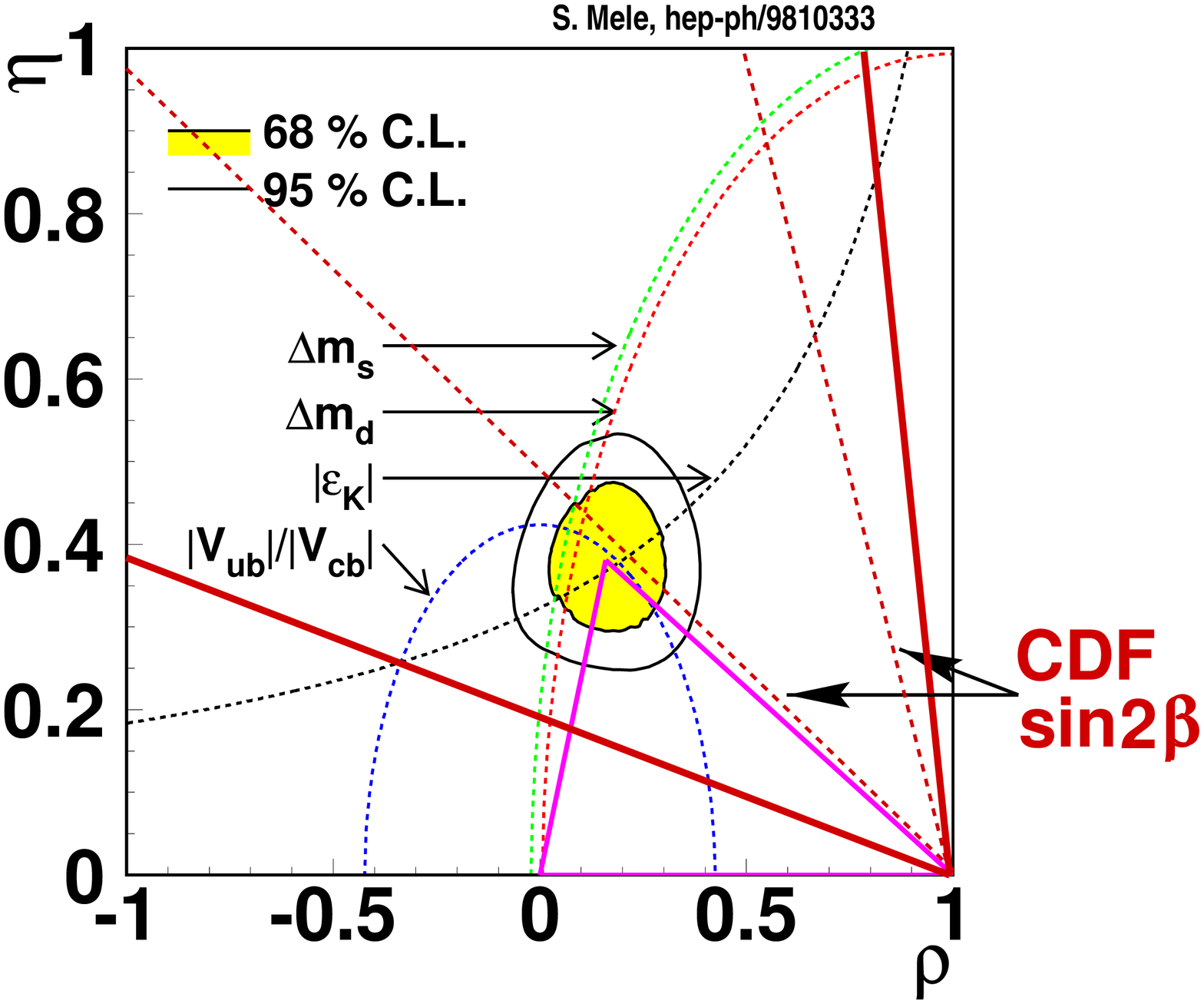}}
   \caption[]
           {\label{fig:sin2beta} Left: The true asymmetry ($\sin (2 \beta )$)
as a function of the proper decay length for the $B \rightarrow
J/\psi \, K^0_{\rm S}$ events. The data points shown have the
effective dilution combined for single and double tagged events
after background subtraction.  
The non-SVX data sample is represented as a single point.
The curves are from the full fit
using both the SVX and non-SVX data.
Right: The measurement of $\sin (2 \beta )$
translates into two results (dotted lines) and one sigma bound (solid
lines) on $\beta$ which is displayed on the $\rho - \eta$ plane. }
\end{figure}

The result
obtained, $\sin (2 \beta ) = 0.79 {+0.41 \atop -0.44}
( {\rm stat + syst })$, provides the
best direct indication so far that $CP$ invariance is violated in the $b$ quark
system. This result is consistent with the Standard (CKM) Model
expectation for a large
positive asymmetry.
Indirect constraints from measurements of other CKM related
quantities suggest that $\sin (2 \beta)$ is large and positive: $\sin (2
\beta) = 0.75 \pm 0.09$.~\cite{ref:mele}  
Fig.~\ref{fig:sin2beta} displays the
result on $\beta$ on the $\rho - \eta$ plane (following
Wolfenstein's~\cite{ref:lincoln}
parametrization of the CKM quark mixing matrix.)
It is noted that the expected sign
of the asymmetry depends on the relative sign of hadronic matrix elements
governing mixing in the $K^0 - \overline{K^0}$ and $B^0 - \overline{B^0}$
systems.~\cite{ref:kayser}  

Extrapolating to the anticipated luminosity
of 2 fb$^{-1}$ in Run II and assuming no improvements in the effective
tagging efficiency, an uncertainty on $\sin (2 \beta)$ of $\sim 0.08$
is expected.\cite{ref:tdr}

\section{Measurement of Polarization Amplitudes in
$B^0 \rightarrow J/\psi \, K^{\ast 0}$ and $B_s^0 \rightarrow J/\psi \,
\phi$ Decays}

The decays $B^0 \rightarrow J/\psi \, K^{\ast 0}$ and $B_s^0 \rightarrow J/\psi \,
\phi$ are pseudoscalar to vector-vector decays and in
principle have three decay amplitudes which can be determined by
studying the angular distributions of the final state particles.
These decays can have orbital angular momenta between the
$J/\psi$ and $K^{\ast} $ (or $\phi$) of 0, 1, or 2,
and three matrix elements are needed to describe the transitions to these
three eigenstates of the $J/\psi$ $K^{\ast}$ (or $\phi$) system. 
A very useful basis for this description 
is the transversity basis.~\cite{ref:transversity}
In this basis one matrix element, $A_\perp$, corresponds to the parity
odd, $L = 1$  (P wave), amplitude, and two matrix elements $A_0$
and $A_\parallel$ are combinations of the parity even, $L = 0$ and $L = 2$
(S and D wave), amplitudes. Also, $|A_0|^2$ is
equal to the longitudinal
polarization fraction, $\Gamma_L /\Gamma$, as is commonly defined
in the helicity
basis.~\cite{ref:VV}

A determination of the longitudinal polarization is relevant
to testing the limitations of theoretical predictions which follow
from the factorization hypothesis.  
The
factorization hypothesis
assumes that the weak decay amplitude can be described as
the product of two independent (hadronic) currents.  
For these decays the factorization ansatz
treats the $J/\psi$ as a current independent of the $B
\rightarrow K^{\ast}$ ($\phi$) current. One assumes the decay matrix elements
factorize naturally into short and long distance (weak and strong)
processes which do not interfere with each other. This implies that the
matrix elements of the decay be relatively real. The observation of
nontrivial phases between the matrix element implies final state interactions
(though the absence of nontrivial phases need not rule out the presence
of final state interactions).


A measurement of the parity odd amplitude, $A_\perp$, is of interest
from the point of view of studies of $CP$ invariance.
As discussed earlier, the decay mode $B^0 \rightarrow J/\psi K_{\rm S}^0$
is useful for a determination of $\sin (2 \beta)$.
This is due to the fact that the final state is a $CP$
eigenstate and one weak amplitude contributes to the decay.  
The decay $B^0 \rightarrow J/\psi~K^{\ast 0}$ can also be of use, when the
final state is a $CP$ eigenstate ({\em e.g.} $K^{\ast 0} \rightarrow
K^0_{\rm S}~\pi^0$).  
A measurement of $\beta$ is most readily
extracted if one or the other parity
amplitude dominates the decay, otherwise the
asymmetry in the decay rates is diluted.  The situation holds
as well for the decay $B_s^0 \rightarrow J/\psi \phi$
which is expected to have a very small $CP$ decay rate asymmetry in
the Standard (CKM) Model.

Finally,
one of the objectives of studying $B^0_s$~meson decays is to determine
the properties of the $B^0_{s,H}$ and $B^0_{s,L}$ states. Since they
are very nearly $CP$ eigenstates, they will decay with distinct angular
distributions. This can improve the sensitivity of a
lifetime difference measurement by adding information beyond
the decay time distribution alone.\cite{ref:deltagamma}

\begin{figure}
   \centerline{\epsfxsize 3.2 truein \epsfbox{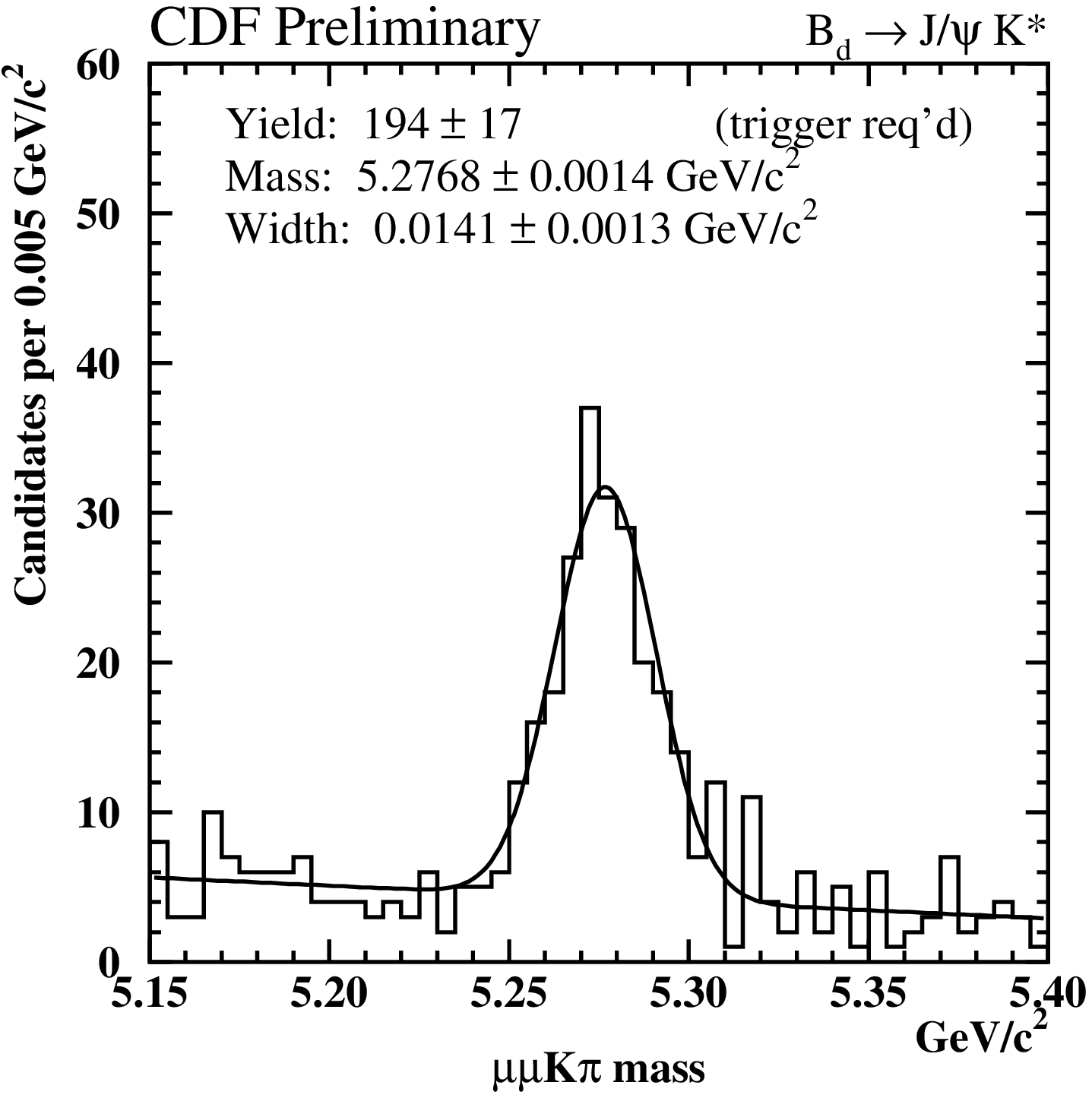}
               \hspace*{0.5cm}
               \epsfxsize 3.2 truein \epsfbox{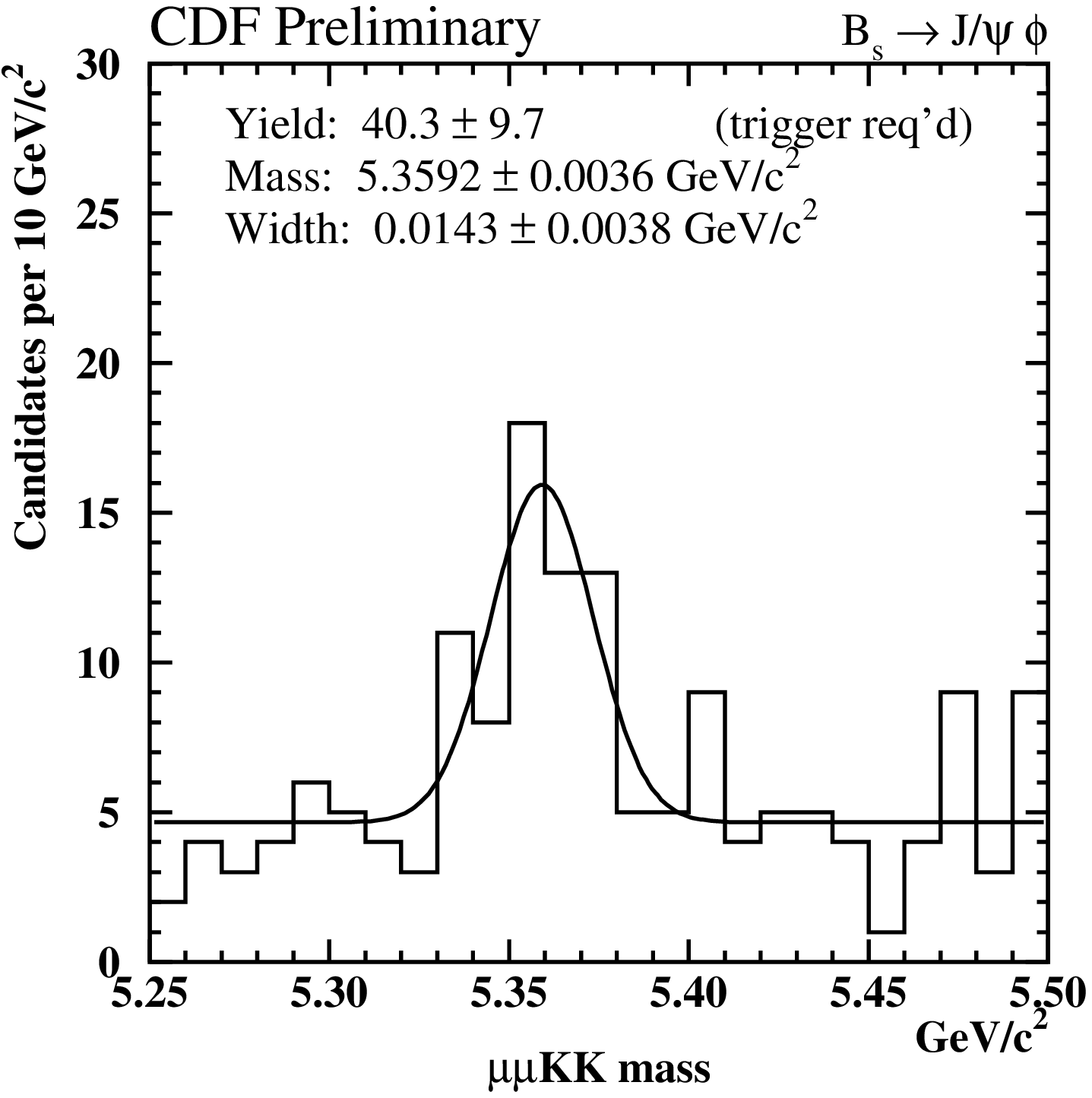}}
   \caption[]
           {\label{fig:vvmass} Mass distributions for $B_d$ (left)
            and $B_s^0$ (right) candidates used in the polarization analysis.}
\end{figure}

The events used in this analysis are selected from a sample
in which the muons from the $J/\psi$ decay satisfy
dimuon
triggers employed during Run~Ib (90 pb$^{-1}$).  The event selection criteria
are similar in spirit to the $J/\psi K_{\rm S}^0$ analysis discussed
above.  The main
differences are the requirements for $p_T > 2.0$ ($1.5$) GeV/$c$ for
the $K^{\ast}$ ($\phi$) and $p_T > 6.0$ ($4.5$) GeV/$c$ for the $B^0$ ($B_s^0$)
candidates.  Also two of the four charged tracks in each candidate are
required to be well measured in the SVX and
a minimum proper decay length of
100 (50) $\mu$m is required for $B^0$ ($B_s^0$) candidates.
In principle, this can bias the angular distribution for the $B_s^0$
since the mass eigenstates are approximately $CP$ eigenstates
and can have different lifetimes.
The mass distributions of the $B$
candidates are shown in Fig.~\ref{fig:vvmass}.

The decay angular distribution has the following form, expressed
in terms of the decay angles of the decay products of the vector
mesons:~\cite{ref:transversity}
\begin{eqnarray}
   \Omega_{\rm Trn} \propto
      &   & \rule{0mm}{0.6cm} 2 \cos^2 \Theta_{K^{\ast}} \left( 1 -
         \sin^2 \Theta_{\rm T} \cos^2 \Phi_{\rm T} \right) |A_0|^2
       + \rule{0mm}{0.6cm} \sin^2 \Theta_{K^{\ast}} \left( 1 - \sin^2
         \Theta_{\rm T}
         \sin^2 \Phi_{\rm T} \right) |A_\parallel|^2 \nonumber \\
      & + & \rule{0mm}{0.6cm}\sin^2 \Theta_{K^{\ast}} \sin^2
         \Theta_{\rm T} |A_\perp|^2
       + \rule{0mm}{0.6cm} \frac{1}{\sqrt{2}}\sin 2\Theta_{K^{\ast}}
         \sin^2 \Theta_{\rm T} \sin 2\Phi_{\rm T}
         \,{\rm Re}(A_0^*A_\parallel) \nonumber \\
      & \mp &
         \rule{0mm}{0.6cm} \sin^2 \Theta_{K^{\ast}}
         \sin 2\Theta_{\rm T} \sin \Phi_{\rm T}
         \,{\rm Im}(A_\parallel^*A_\perp)
      \pm \rule{0mm}{0.6cm}
         \frac{1}{\sqrt{2}}\sin 2\Theta_{K^{\ast}} \sin 2\Theta_{\rm T}
         \cos \Phi_{\rm T} \,{\rm Im}(A_0^*A_\perp) \nonumber
\end{eqnarray}
Note that the last
two terms have opposite signs for the decay of a $\bar{B}$ as
compared with a $B$. The $B^0$ and $\overline{B^0}$ decays are flavor tagged
by the charge of the $K$ meson, but the $B_s^0$ and $\overline{B}_s$ are not
distinguishable by their final state particles. Hence, for $B_s^0$
decays
information about the phase of $A_\perp$ is lost.

\begin{figure}
   \centerline{\epsfxsize 3.2 truein \epsfbox{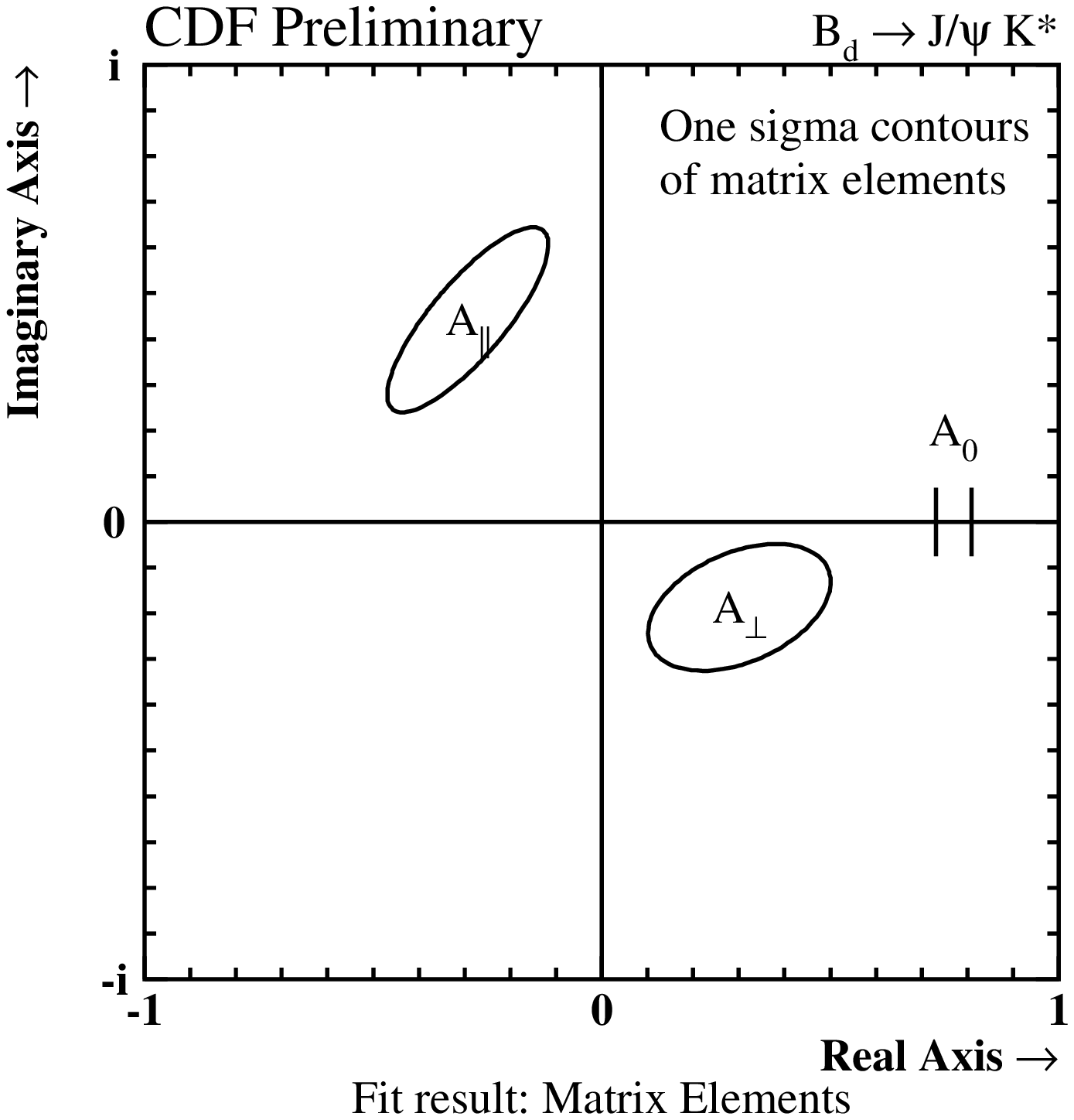}
               \hspace*{2mm}
               \epsfxsize 3.2 truein \epsfbox{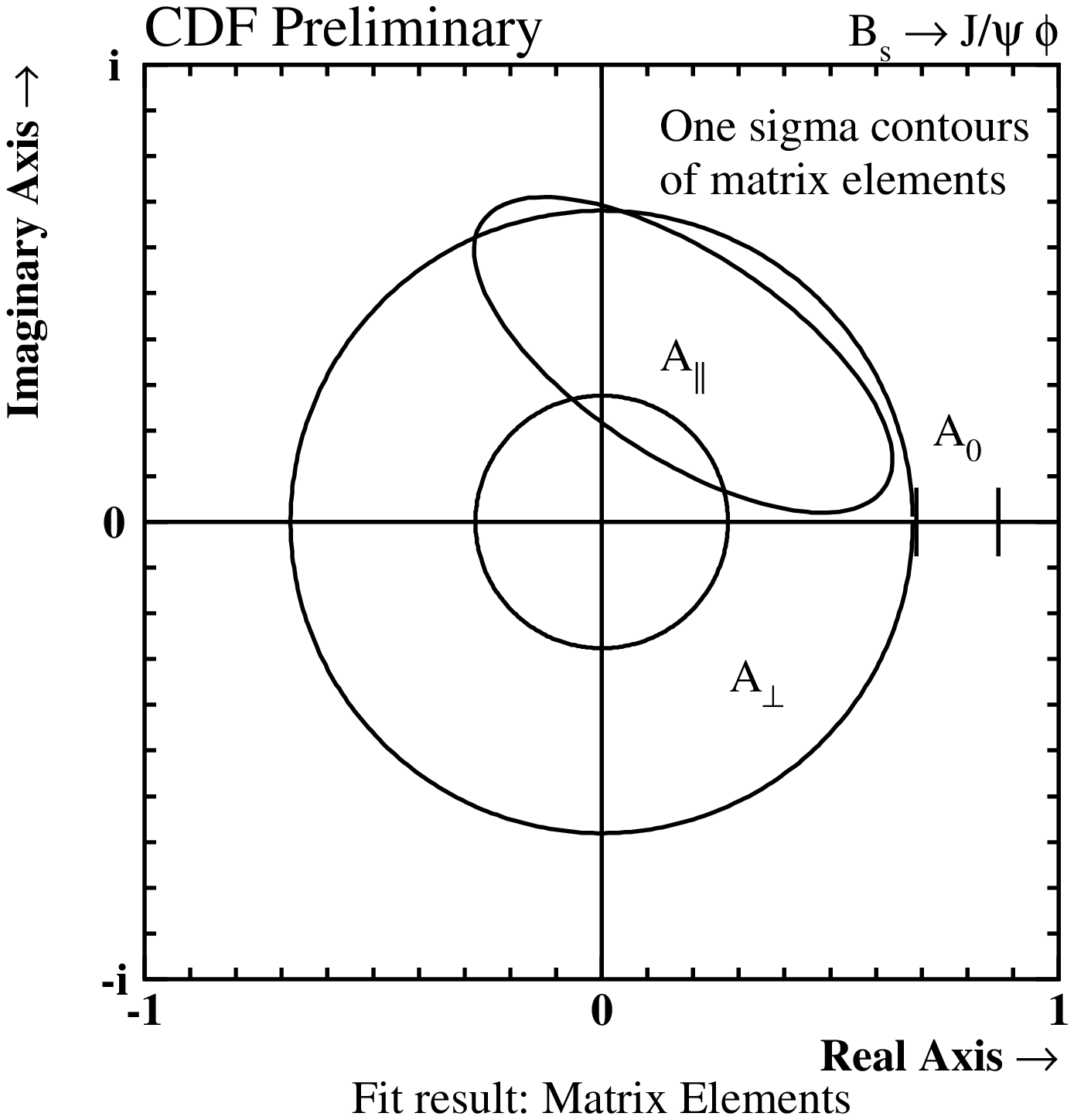}
}
   \caption[]
           {\label{fig:contours} One
            sigma contours of the fit for the decay amplitudes
in the $B^0 \rightarrow J/\psi K^{\ast 0}$ and
$B_s^0 \rightarrow J/\psi \phi$ decay modes; the longitudinal component
($A_0$) is taken to be real.
}
\end{figure}

The decay matrix elements are extracted
from a likelihood fit
with care taken into 
account for the detector acceptance and residual
backgrounds.~\cite{ref:morepol}
Fig.~\ref{fig:contours} shows one sigma
contours for the extracted decay matrix elements.  For the $B^0$
decay:
\begin{equation}
   \begin{array}[c]{rcl}
      A_0         & = & 0.770 \pm 0.039 \pm 0.012 \\
      A_\parallel & = & (0.530 \pm 0.106 \pm 0.034) e^{(2.16 \pm 0.46
                        \pm 0.10) i} \\
      A_\perp     & = & (0.355 \pm 0.156 \pm 0.039) e^{(-0.56 \pm 0.53
                        \pm 0.12) i}
   \end{array}
\end{equation}
\noindent
and
\begin{eqnarray}
   |A_0|^2 = \frac{\Gamma_L}{\Gamma}  & = &
     0.593^{+0.059}_{-0.061} \pm 0.018   \\
    |A_\perp|^2 = \frac{\Gamma_\perp}{\Gamma}
      & = &  0.126^{+0.121}_{-0.093} \pm 0.028 \nonumber
\end{eqnarray}
\noindent
and for the $B_s^0$:
\begin{equation}
   \begin{array}[c]{rcl}
      A_0         & = & 0.778 \pm 0.090 \pm 0.012 \\
      A_\parallel & = & (0.407 \pm 0.232 \pm 0.034) e^{(1.12 \pm 1.29
                         \pm 0.11) i} \\
      |A_\perp|   & = & 0.478 \pm 0.202 \pm 0.040
   \end{array}
\end{equation}
\noindent
and
\begin{eqnarray}
   |A_0|^2 = \frac{\Gamma_L}{\Gamma}  & = &  0.606 \pm 0.139 \pm 0.018 \\
   |A_\perp|^2 = \frac{\Gamma_\perp}{\Gamma}
      & = &  0.229 \pm 0.188 \pm 0.038 \nonumber
\end{eqnarray}

\noindent
The $B^0$ results are of comparable sensitivity to the results from
CLEO;~\cite{ref:cleo}
comparable magnitudes are obtained for the three matrix elements,
but different central values for the
phases. The phases observed in the CDF analysis leave open the possibility
of
non-trivial final state interactions in the decay.
The CDF Run~Ib result for the longitudinal polarization
is in good agreement with the CDF Run~Ia result.~\cite{ref:randy}
This is an important result for tests of
factorization,~\cite{ref:factor} especially
when considered along with the observed
ratio of  branching ratios,  R$={\cal B} (B \rightarrow J/\psi K^{\ast}
)/{\cal B} (B \rightarrow J/\psi K)$.~\cite{ref:cleo,ref:br}

The $B_s^0$ results are the first and only ones available for a full angular
analysis.  Again the Run~Ib result for the longitudinal polarization
is in agreement with that from Run~Ia.~\cite{ref:randy}
Comparison of the $B^0$ and $B_s^0$ results indicates that $SU(3)_{\rm
flavor}$ is a valid approximation. 
The decays are dominated by the parity even amplitudes but a non-trivial
parity odd component is not yet excluded.

\section{Conclusions}

New results have been presented for a measurement of $\sin (2 \beta)$
from $B^0 \rightarrow J/\psi K_{\rm S}^0$ decays and for full polarization
analyses of the
 decays $B^0 \rightarrow J/\psi \, K^{\ast 0}$ and $B_s^0 \rightarrow J/\psi \,
\phi$.  Besides being of interest in themselves, these results whet
the appetite for the richness of the CDF program on $b$ physics during
Run~II, scheduled to start in the summer of 2000.

\section*{Acknowledgements} This work would not be possible without
the vital contributions of the staff at Fermilab and all the
members and
technical staff of the collaborating institutions, and support from
the funding agencies.  The author would like to thank H.~Lipkin for
bringing the transversity basis to our attention and acknowledge
illuminating correspondance and discussions with I.~Dunietz and J.~Rosner.

\end{document}